\documentclass{iopart}
\usepackage{iopams,amsfonts,amssymb,amsthm}
\expandafter\let\csname equation*\endcsname\relax		
\expandafter\let\csname endequation*\endcsname\relax	
\usepackage{amsmath}

\usepackage[pdftex]{graphicx}       
\graphicspath{{figs/}}
\usepackage{txfonts}        
\usepackage{mathrsfs}	
\usepackage{orcidlink}	
\usepackage{layout}

\newcommand{\Ta}{T_\text{a}}

\newcommand{\orderparameter}[1]{\frac{\langle #1 \rangle}{n}}
\newcommand{\orderparameterinline}[1]{\langle #1 \rangle / n}

\DeclareMathOperator{\var}{var }
\usepackage{hyperref} 
\hypersetup{colorlinks,citecolor=blue,filecolor=blue,linkcolor=blue,urlcolor=blue}

\begin{document}

\title{Self-avoiding walks pulled at an angle}

\author{
	C J Bradly$^1$\footnote{Author to whom any correspondence should be addressed.}\orcidlink{0000-0002-5413-777X},
	N R Beaton$^1$\orcidlink{0000-0001-8220-3917} and 
	A L Owczarek\orcidlink{0000-0001-8919-3424}}
\address{
	$^1$School of Mathematics and Statistics, University of Melbourne, Victoria 3010, Australia}
\ead{
	\href{mailto:chris.bradly@unimelb.edu.au}{chris.bradly@unimelb.edu.au}
	}

\vspace{10pt}
\begin{indented}
\item[]\today
\end{indented}

\begin{abstract}
We investigate polymers pulled away from an interacting surface, where the force is applied to the untethered endpoint and at an angle $\theta$ to the surface.
We use the canonical self-avoiding walk model of polymers and obtain the phase diagram of the model using Monte Carlo simulations for a range of angles, temperatures and force magnitudes.
The phase diagram of the model displays re-entrance at low temperatures for three-dimensional walks when the pulling is more vertical than horizontal.
Our results agree with various exactly solvable lattice models that have been previously studied.
\end{abstract}



\section{Introduction}
\label{sec:Intro}


The manipulation of individual long chain molecules that are attached to a substrate is possible with atomic force microscopy \cite{Haupt1999,Bemis1999,Zhang2003}.
Experimental techniques have advanced to provide more fine-grained control over the AFM tip \cite{Kuehner2006,Grebikova2017}.
A recent experiment \cite{Grebikova2018} used atomic force microscopy to measure the desorption force as a function of the angle to the surface.
The experimental results were found to be universal across several polymer-substrate systems and had qualitative agreement with the predictions of partially directed walk models ~\cite{Osborn2010,Orlandini2010}.

The canonical lattice model that best matches the configurational space of real polymers is the self-avoiding walk (SAW).
The case where SAWs are desorbed due to a vertical force applied at the endpoint has been studied \cite{Guttmann2014,Rensburg2013,Mishra2005,Krawczyk2004}.
Some variation on how the force is applied have also been studied, for example if the force is applied at the midpoint \cite{Rensburg2017} or an arbitrary interior point \cite{Bradly2019d}.
However, the self-avoiding walk model has not been studied in the context of a force applied at an angle.

Another lattice model, that of partially directed self-avoiding walks (PDWs) subject to a pulling force applied at an angle, was studied some years ago.
PDWs are a subset of SAWs defined by disallowing steps in one or more directions, and the directedness allows this model to be solved exactly in many scenarios.
The two-dimensional case of a PDW pulled at an angle was solved by Refs.~\cite{Osborn2010,Orlandini2010}, both finding the complete phase diagram.
The direction of the pulling force has a similar effect as anisotropic stiffness in the location of the desorption transition \cite{Tabbara2012}.
Three-dimensional models were also solved in Ref.~\cite{Osborn2010}, where there is flexibility in both the definition of a partially directed walk (i.e.~which directions are forbidden) as well as the direction of the applied force with respect to its projection in the impermeable surface.
One of these three-dimensional models can be identified as also including an inhomogeneous surface \cite{Iliev2012}.

\begin{figure}[t!]
	\centering
	\includegraphics[width=0.6\textwidth]{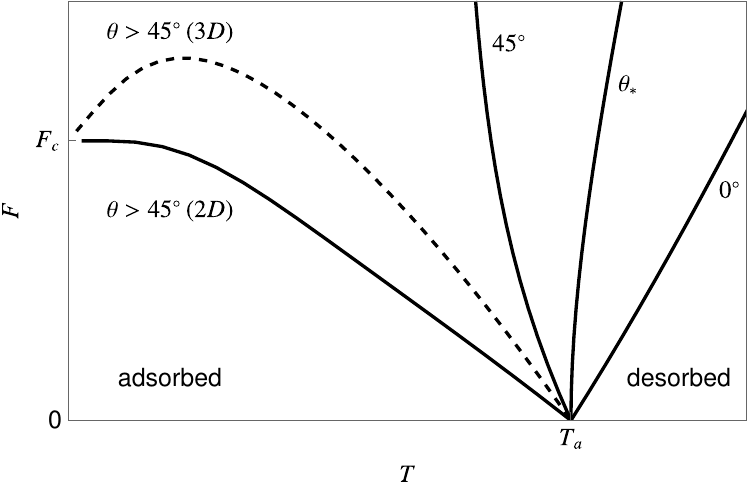}
	\caption{The schematic phase diagram of pulled walks based on the solution to partially-directed walk models.
	The adsorbed phase is always on the left and the desorbed phase is always on the right but the boundary between them depends on the angle of the applied force.
	}
	\label{fig:Schematic}
\end{figure}

The known features of pulled lattice polymers are illustrated in a schematic phase diagram in \fref{fig:Schematic} based on the PDW solutions from Ref.~\cite{Osborn2010}.
When there is no applied force there is a thermal transition at the adsorption point $\Ta$.
When the force $F$ is applied the phase boundary between adsorbed and desorbed phases depends strongly on the angle $\theta$ between the force and the interacting surface.
When the force is applied horizontally (or near-horizontally) then at low temperatures the model is always adsorbed regardless of $F$ and at high-temperatures there is force-induced \emph{adsorption}, by which we mean that as the force is increased the transition is from the desorbed phase to the adsorbed phase.
Conversely, when the force is applied vertically (or near-vertically) then at high temperatures the model is always desorbed for all $F$ and at low-temperatures there is force-induced \emph{desorption}.
The dividing point between these two behaviours is simply $\theta = 45^\circ$ where the horizontal and vertical components are balanced, and this is the largest angle for which the phase boundary does not exist at $T = 0$.
For larger angles, there is a zero-temperature critical force $F_c$, which does depend on $\theta$.

The other interesting feature of the phase diagram is the phenomenon of phase re-entrance.
We will refer to a phase being $X$--re-entrant if varying a parameter $X$ (with all other parameters fixed) causes a transition out of the phase and then back into the phase at two different values of $X$.
The clearest example in \fref{fig:Schematic} is the low temperature region where the desorbed phase is $T$--re-entrant in three-dimensions when the pulling force is near-vertical (dashed line).
This is due to the non-zero entropy of the adsorbed phase of the three-dimensional model. (On the other hand, the desorbed phase is never $T$--re-entrant for the two-dimensional model.)
The range of forces for which the desorbed phase is $T$--re-entrant shrinks as $\theta$ is reduced from vertical pulling but is still present for angles less than $90^\circ$.
A less obvious example is the low force region where the adsorbed phase is $F$--re-entrant for a certain range of angles.
The solution to the PDW model indicates that there is some angle $\theta_*$ for which the boundary is vertical in the $T$--$F$ plane as $F \to 0$.
This means that for angles $\theta_* < \theta < 45^\circ$ and fixed temperature just below $\Ta$ the adsorbed phase is $F$--re-entrant; one can increase the force from zero and encounter two transitions from adsorbed to desorbed and back to adsorbed phase.

While SAWs are the canonical lattice model of polymers the addition of a pulling force with a horizontal component imbues it with a directedness and so we may expect similarities with partially directed walks \cite{Mishra2005,Owczarek2010}.
Our goal in this paper is to investigate whether the features of the partially directed model are present in the self-avoiding model.
When the pulling force is strictly vertical we know that there is $T$--re-entrance at low temperatures in three-dimensional SAWs \cite{Krawczyk2004} but other angles are unknown.
In \sref{sec:Model} we describe the self-avoiding walk model and present the phase diagram based on simulation results in \sref{sec:Results}.
We discuss several features of the SAW phase diagram in \sref{sec:Discussion} including relevant scaling arguments and comparison to the partially directed model.

\begin{figure}[t!]
	\centering
	\includegraphics[width=0.5\textwidth]{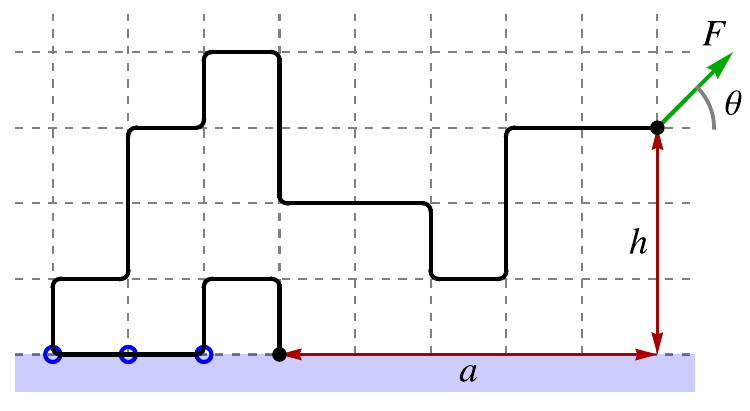}
	\caption{A self-avoiding walk pulled at an angle $\theta$ by a force $F$. There are $m=3$ vertices in the surface and the endpoint is at $(a,h) = (5,3)$.}
	\label{fig:Model}
\end{figure}

\section{Model of pulled walks}
\label{sec:Model}

The polymers are modeled as self-avoiding walks on the square ($d = 2$) or simple cubic ($d = 3$) lattice $\mathbb{Z}_d$.
The walks are restricted to the upper half-space so that there is an impermeable interacting surface at the $x$-axis for $d = 2$ or the $x$-$y$ plane for $d = 3$.
The first vertex of the walk is fixed at the origin and an interaction energy $\epsilon < 0$ is attributed to all other vertices that lie in the surface.
The untethered endpoint is pulled by a force with magnitude $F$, applied at an angle $\theta$ to the positive $x$-axis.
The walks are thus enumerated by the number of contacts with the surface $m$, the height of the pulled endpoint $h$ and the projection of the endpoint down to the $x$-axis $a$; see \fref{fig:Model}.

The number of walks of length $n$ steps is denoted $c_{nmah}$, leading to the partition function
\begin{equation}
	Z_n(T, F, \theta) = \sum_{m,a,h} c_{nmah} \kappa^m \lambda^a \tau^h,
	\label{eq:PartitionFunction}
\end{equation}
where the walks are weighted by Boltzmann weights
\begin{equation}
	\kappa = e^{-\epsilon / k_B T}, \quad \lambda = e^{F b \cos \theta / k_B T}, \quad \tau = e^{F b \sin \theta / k_B T}
	\label{eq:BoltzmannToPhysical}
\end{equation}
These weights are expressed in terms of physical parameters $(T, F, \theta)$ which will be more useful for understanding the phase diagram.
We will use lattice units where the lattice spacing is $b = -\epsilon = k_B = 1$.
To determine the phase diagram we consider the average adsorbed fraction 
\begin{equation}
	\orderparameter{m} = \frac{\sum_{m,a,h} m \, c_{nmah} \kappa^m \lambda^a \tau^h}{n \, Z_n},
	\label{eq:OrderParameter}
\end{equation}
The average quantities $\orderparameterinline{a}$ and $\orderparameterinline{h}$ are calculated similarly.

The model \eref{eq:PartitionFunction} is defined in terms of Boltzmann weights and the Monte Carlo simulations work in this formulation, but for pulled walks it will be insightful to consider the phase diagram in terms of physical parameters $T$, $F$ and $\theta$.
The adsorbed fraction $\orderparameterinline{m}$ is the standard order parameter since it is conjugate to the interaction with the surface.
Generally there are two phases: an adsorbed phase where $\orderparameterinline{m} > 0$ (and in fact $\orderparameterinline{m} \to 1$ as $T \to 0$) and a desorbed phase where $\orderparameterinline{m} = 0$.

The phase diagram can also be determined by considering $\orderparameterinline{a}$ and $\orderparameterinline{h}$ together. 
The vertical extent of the endpoint $\orderparameterinline{h}$ is always zero in the adsorbed phase and non-zero in the desorbed (and stretched) phase although its finite value depends on the temperature and force, except when the force is zero. We note here that the desorbed phase is of a slightly different character when the vertical force is zero with $\langle h \rangle \sim n^\nu$, where $\nu$ is the metric exponent for free walks. 
The behaviour of the horizontal extent $\orderparameterinline{a}$ is roughly inverse to the vertical extent; it is non-zero and temperature dependent in the adsorbed phase and in the desorbed phase it is constant, but not necessarily zero, depending on the angle $\theta$.


More details will be discussed with reference to the simulation results in \sref{sec:Results} but some features of the phase diagram are known based on similar models.
There is a first-order phase transition when the angle is non-zero at a transition point between the two phases but the location and nature of this point depends on $F$ and $\theta$.
At zero force there cannot be any angle dependence and the critical point is the second-order adsorption transition at a known temperature $\Ta$ \cite{Grassberger2005,Bradly2018}.
When the force is vertical ($\theta = 90^\circ$) there is force-induced desorption at low temperatures, and at high temperatures there is no adsorbed phase regardless of force magnitude \cite{Krawczyk2004}.
Conversely, when the force is horizontal ($\theta = 0^\circ$) there is force-induced \emph{adsorption} at high temperatures and no desorbed phase at low temperatures for any force magnitude.

\subsection{Simulation of SAWs}
\label{sec:Simulation}

The self-avoiding walk model is analysed using Monte Carlo simulation, based on the parallelised flatPERM algorithm \cite{Prellberg2004,Campbell2020}.
This algorithm samples walks by starting at the origin and growing the endpoint using a pruning and enrichment strategy up to a set maximum length.
Typically with flatPERM, at each step the microcanonical parameters (in our case $m$, $a$ and $h$) of the sample are determined and the cumulative Rosenbluth weight of the walk is recorded in a histogram $W_{nmah}$ which approximates the counts $c_{nmah}$.
The algorithm is athermal in that the Boltzmann weights are added after the simulation to calculate thermodynamic quantities, derived from the partition function \eref{eq:PartitionFunction}.
In the present model, the histogram would be four-dimensional, which is prohibitively expensive to obtain samples for useful lengths.
The typical resolution for models with multiple parameters is to fix one of the Boltzmann weights for the whole simulation, effectively flattening a dimension of the histogram, equivalent to performing one of the sums of \eref{eq:PartitionFunction} within the simulation.
However, for our current model, the phase diagram is such that we wish to sweep over physical parameters $T$, $F$ and $\theta$, which cannot be easily disentangled from the Boltzmann weights weights $\kappa$, $\lambda$ and $\tau$.
Thus we need to perform an independent simulation for each set of values $(T,F,\theta)$.

The output of each simulation is a set of one-dimensional histograms $W_n(T,F,\theta; Q)$ (indexed by $n$), whose entries are the \emph{sample} averages of the weighted quantity $Q \, w \, \kappa^m \lambda^a \tau^h$ for samples of length $n$.
The cumulative Rosenbluth weight $w$ of a sample approximates $c_{nmah}$ and $Q$ is a microcanonical quantity.
The quantities of interest are all zeroth-, first- and second-order moments of $m$, $a$ and $h$ in order to approximate all relevant thermodynamic quantities.

The maximum length achievable in any simulation depends on the parameters $(T,F,\theta)$.
In particular, the minimum value of $T$ determines how large $\kappa^n$ is, which cannot exceed the maximum value that can be handled with floating-point arithmetic (we use quadruple precision types).
Typically we ran simulations sampling walks up to length $n = 500$ using 4 parallel threads, each running $10^6$ flatPERM iterations.
As an example, the phase diagrams in \fref{fig:PhaseBoundaries} has $37\times37\times7$ points in the $(T,F,\theta)$ space, for a total of approximately 50,000 CPU-hours of simulation runtime for each of $d=2$ and $3$.
To obtain data for very small $T$ we ran some additional simulations up to only $n = 128$, which is still long enough to resolve the phase diagram.

\begin{figure}[t!]
	\centering
	\includegraphics[width=\textwidth]{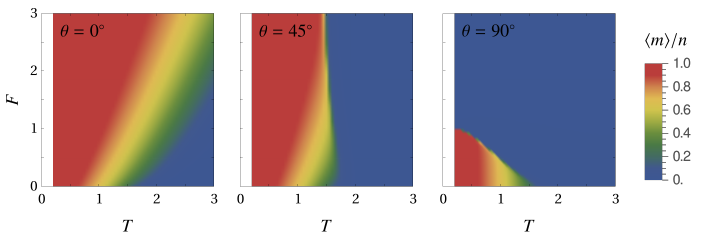}
	\includegraphics[width=\textwidth]{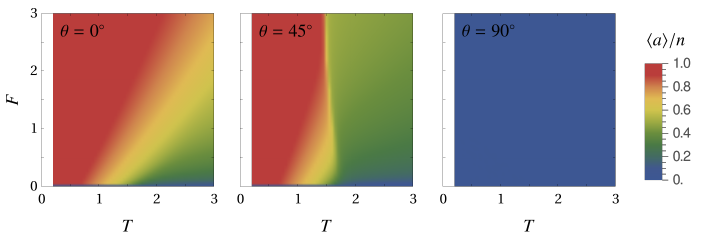}
	\includegraphics[width=\textwidth]{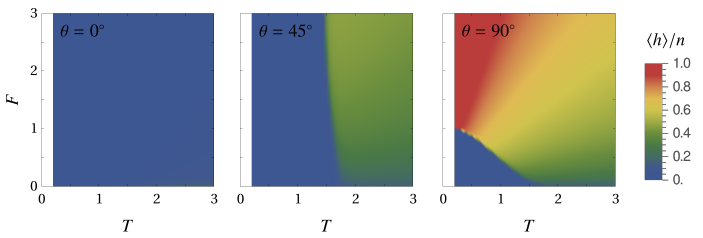}
	\caption{The order parameters for $d = 2$ SAWs pulled at an angle $\theta = 0^\circ, 45^\circ, 90^\circ$ (left to right).
	From top to bottom are the adsorbed fraction $\orderparameterinline{m}$, average horizontal position of the endpoint $\orderparameterinline{a}$ and average vertical position of the endpoint $\orderparameterinline{h}$.
	For all angles, the desorbed phase lies on the right of the boundary, and the adsorbed phase lies on the left side.}
	\label{fig:OrderParameters2D}
\end{figure}

\section{Phase diagram}
\label{sec:Results}

To see the phase diagram, we use the simulation data to calculate the order parameter as
\begin{equation}
	\orderparameter{m} \approx \frac{W_n(T,F,\theta; m)}{n \, W_n(T,F,\theta; 1)},
	\label{eq:OrderParameterSimulation}
\end{equation}
and similar for $\orderparameterinline{a}$ and $\orderparameterinline{h}$.
In \fref{fig:OrderParameters2D} we show the order parameters for $d = 2$ SAWs and for $\theta = 0^\circ, 45^\circ, 90^\circ$ (left to right).
The existence of two phases is clear and the boundary depends strongly on $\theta$.
In all cases the adsorbed phase is on the left part of the region indicated by $\orderparameterinline{m} > 0$ and $\orderparameterinline{h} \approx 0$ (blue).
In this phase $\orderparameterinline{a} > 0$ also except for $\theta = 90^\circ$ where there is no preferred direction to break the symmetry.
Furthermore in the desorbed phase the horizontal component of the force means that $\orderparameterinline{a} > 0$ as well, albeit a smaller value than in the adsorbed phase.
Similarly, the phase boundary is not visible from $\orderparameterinline{h}$ for $\theta = 0^\circ$ since there is no vertical component in that case, and for angles inbetween ($\theta = 45^\circ$) the contrast between phases is not as strong compared to $\orderparameterinline{m}$.
For these reasons we consider $\orderparameterinline{m}$ to be the best order parameter for the model.

\begin{figure}[t!]
	\centering
	\includegraphics[width=\textwidth]{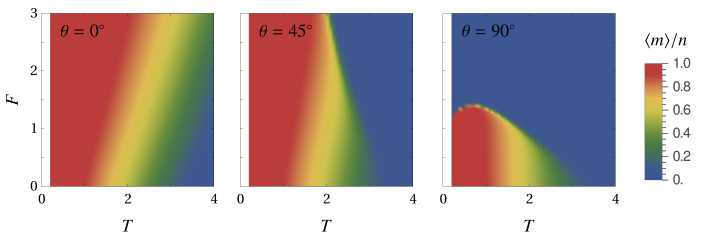}
	\includegraphics[width=\textwidth]{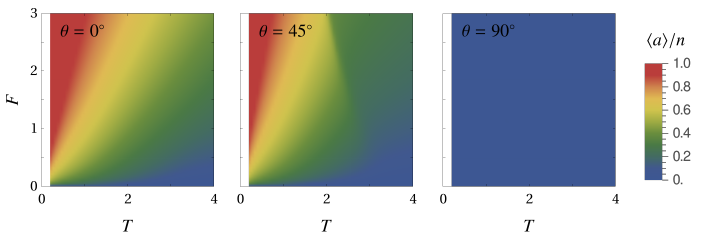}
	\includegraphics[width=\textwidth]{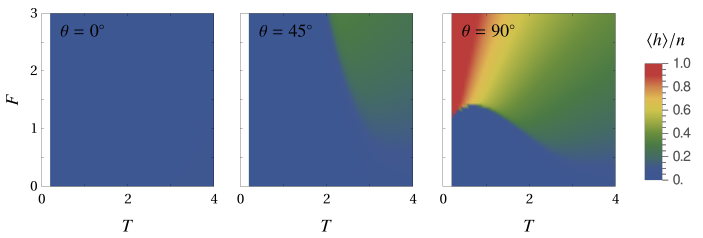}
	\caption{The order parameters for $d = 3$ SAWs, with same parameters as \fref{fig:OrderParameters2D}.}
	\label{fig:OrderParameters3D}
\end{figure}

In \fref{fig:OrderParameters3D} we show the order parameters for $d = 3$ SAWs, for the same parameters as \fref{fig:OrderParameters2D}, except for a slightly larger range in $T$.
The phase diagram is qualitatively similar to the $d = 2$ model, with a few differences.
The main difference is that the adsorbed phase is $T$--re-entrant at low $T$ for $\theta = 90^\circ$ at fixed values of $F$ just above 1.
Otherwise, $\orderparameterinline{a}$ and $\orderparameterinline{h}$ are even less of a reliable indicator of the phase structure than the $d = 2$ case since now the extra dimension is perpendicular to the applied force.

While the phases are clear from $\orderparameterinline{m}$, it is easier to see the boundaries more clearly by calculating a covariance, which generalises the specific heat.
Thus we also consider the Hessian matrix of second derivatives of the reduced free energy $A_n = - n^{-1} \log Z_n$, with respect to $\kappa$, $\lambda$, $\tau$.
\begin{equation}
	H_n = \left(\begin{matrix}
		\partial_\kappa^2 A_n & \partial_\kappa \partial_\lambda A_n & \partial_\kappa \partial_\tau A_n \\
		\partial_\lambda \partial_\kappa A_n & \partial_\lambda^2 A_n & \partial_\lambda \partial_\tau A_n \\
		\partial_\tau \partial_\kappa A_n & \partial_\tau \partial_\lambda A_n & \partial_\tau^2 A_n
	\end{matrix}\right).
	\label{eq:Hessian}
\end{equation}
These derivatives are estimated using simulation data for the first and second moments of $m$, $x$ and $h$.
We denote the largest eigenvalue of $H_n$ by $\chi$ and large $\chi$ is associated with a phase transition.
Because the adsorbed fraction is the best order parameter in this model it would also be sufficient to simply look for peaks in $\var(m)$ to indicate the locations of the phase transition.

\begin{figure}[t!]
	\centering
	\includegraphics[width=\textwidth]{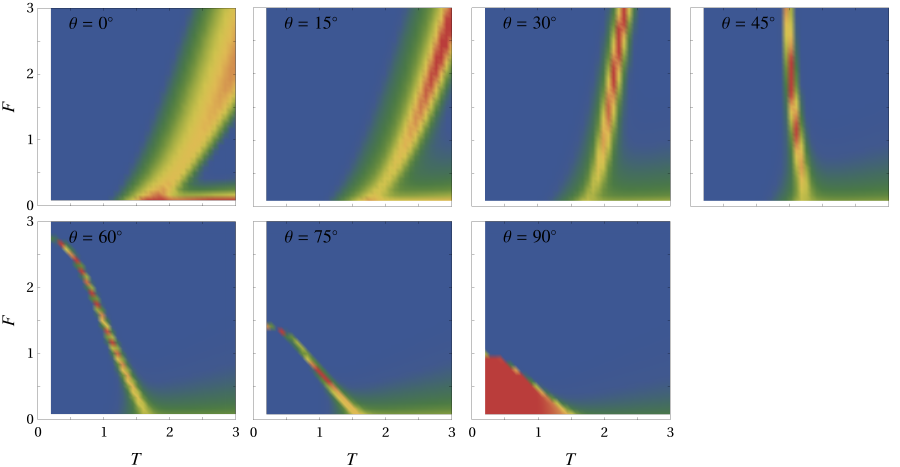}

	\vspace{0.5cm}
	\includegraphics[width=\textwidth]{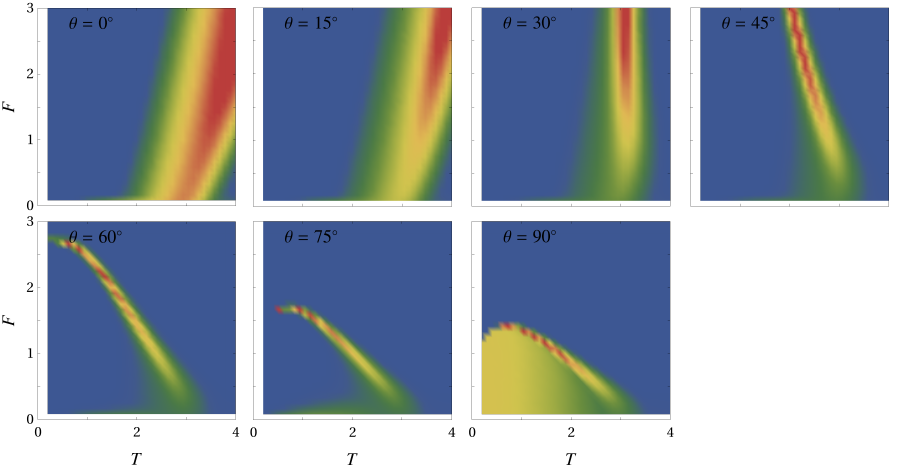}
	\caption{The phase boundaries of the SAW model indicated by regions of high covariance $\log\chi$. The top panel is for $d = 2$ and the bottom panel is for $d = 3$.}
	\label{fig:PhaseBoundaries}
\end{figure}

In \fref{fig:PhaseBoundaries} we indicate the phase boundaries determined from peaks in $\chi$ for (top) $d = 2$ and (bottom) $d = 3$.
In the low-force region all curves meet at a common point independent of $\theta$, which is the adsorption transition.
The present simulations are not sufficient to show the SAW adsorption point accurately, but previous Monte Carlo simulations \cite{Bradly2018} give $\Ta = 1.744(6)$ for $d = 2$ and $\Ta = 3.520(6)$ for $d = 3$.
These values are consistent with the location of the boundaries as $F \to 0$.
The other minor quantitative difference is that for low angles and fixed temperature (above the adsorption point $\Ta$) the critical force inducing adsorption is slightly higher for $d = 2$, presumably because the change in entropy between adsorbed and desorbed phases is higher for $d = 2$ than for $d = 3$.

Another obvious feature of the phase boundaries is that for large $\theta$ the boundaries join the $F$-axis at finite $F$ meaning that there is force-induced desorption at low-temperature.
Furthermore, the desorbed phase is $T$--re-entrant at low-temperature for $d = 3$ when $\theta$ is large but never $T$--re-entrant for $d = 2$.
The region where re-entrance occurs diminishes as $\theta$ decreases from $90^\circ$ and is barely visible for $\theta = 75^\circ$.
For small $\theta$ the low-temperature phase is adsorbed only, regardless of the magnitude of the force.
Conversely, at high temperatures the model is always in the adsorbed phase when the force is near-vertical, regardless of the magnitude $F$.
For low angles, increasing $F$ causes the walk to adsorb.
The distinction between these two behaviours appears to be around $\theta = 45^\circ$, and this will be justified below with comparison to the partially directed walk model.

A note on \fref{fig:PhaseBoundaries}: no scale is given for the value of $\log\chi$ since it is different for each plot in order to highlight where the boundaries are.
We do not investigate finite-size scaling of the height of the peaks since our data is not sufficient for that purpose.
Furthermore, there are some spurious features in the covariance data that do not correspond to real transitions.
For example, in the plot of the phase boundary for $d = 2$ and $\theta = 90^\circ$, it appears that there is no boundary in this case and instead the entire adsorbed phase has high covariance.
This is an artifact of this special case where there is no preferred horizontal component of the force.
This means we have $\langle a \rangle \approx 0$ but the actual samples consist of two equally likely configurations, where the endpoint has horizontal span $a \approx \pm n$.
Thus in this case $\langle a^2 \rangle > 0$ and the covariance is non-zero even in the adsorbed phase.
The same effect is particularly strong in the $d = 2$ case at $F=0$ for small $\theta$  where it makes other features  barely visible and this data has been omitted to make the adsorbed-desorbed boundary visible.
This same phenomenon also appears in the $d = 3$ case for $\theta = 90^\circ$ but with reduced effect because the endpoint has two-dimensional degree of freedom.
In all other cases where $\langle a \rangle \approx 0$ it is because the samples actually have $a \approx 0$.
These issues indicate that $\orderparameterinline{a}$ is problematic as a microcanonical parameter in limiting cases.

\section{Discussion}
\label{sec:Discussion}


\subsection{Scaling near the adsorption point}
\label{sec:AdsorptionScaling}

The phase boundaries all meet at the adsorption transition where $F = 0$ and $T = \Ta$.
In the absence of a force this is a well understood continuous phase transition.
With an arbitrary force acting on walks we suppose a standard scaling ansatz form for the partition function near the critical point
\begin{equation}
	Z_n(T, F) \sim \mu^n \, n^{\gamma - 1} \mathcal{Z}\left( (T - \Ta) \, n^\phi, F \, n^\psi \right),
	\label{eq:PartitionGeneralScaling}
\end{equation}
where $\mathcal{Z}$ is an unknown analytic function and $\mu$ and $\gamma$ are universal, depending only on the phase and dimension and are otherwise independent of $T$ and $F$. 
At the adsorption point they take their critical values $\mu_c$ and $\gamma_c$.
The crossover exponents $\phi$ and $\psi$ in the scaling ansatz behave similarly.
Then we can form the reduced free energy
\begin{equation}
	A_n(T, F) \sim -\frac{1}{n} (\gamma - 1) \log n - \log \mu + \frac{1}{n} \mathcal{A}\left( (T - \Ta) \, n^\phi, F \, n^\psi \right),
	\label{eq:FreeEnergyGeneralScaling}
\end{equation}
where $\mathcal{A}$ is another unknown analytic scaling function.
In the large $n$ limit we see that the connective constant is related to the limiting free energy per step $\mu = e^{-A_\infty}$.
At zero force we can compute the internal energy
\begin{equation}
	u_n(T) = \orderparameter{m} = \frac{\partial A_n}{\partial T} = n^{\phi - 1} \frac{\partial \mathcal{A}}{\partial T}\left( (T - \Ta) \, n^\phi, 0 \right)
\end{equation}
since only the scaling function depends on $T$ and $F$.
This gives a scaling form for the internal energy, and the value of the exponent $\phi$ at the critical point has attracted attention over the years. 
It is now known to have the value $1/2$ except for $d = 3$ where $\phi \approx 0.48$ \cite{Bradly2018,Grassberger2005}.
At the adsorption point $T = \Ta$ we can do the same for force near $F = 0$ and thus we can obtain the end-to-end distance $R$ as 
\begin{equation}
	\frac{\partial A_n}{\partial F} = \frac{R}{n} \sim n^{\psi - 1} \, \frac{\partial \mathcal{A}}{\partial F}\left( 0, F \, n^\psi \right).
	\label{eq:ForceResponse}
\end{equation}
But it is standard that $R = \langle r_n^2 \rangle^{1/2} \sim n^\nu$ scales with metric exponent $\nu$, and thus $\psi = \nu$.

So far we have effectively integrated out the direction of the force and only consider the magnitude $F$, which is valid for the full lattice, which is isotropic in the thermodynamic limit of long chains.
However, if there is a surface to break this symmetry, one might imagine that the direction of the force affects the scaling.
It is not necessary to consider a specific angle, but simply the components of $R$ that are parallel and perpendicular to the surface, namely $R_\parallel$ and $R_\perp$ and their associated exponents $\nu_\parallel$ and $\nu_\perp$, respectively.
We could expand \eref{eq:PartitionGeneralScaling} to include $ F_\parallel$ and $F_\perp$ along with associated crossover exponents $\psi_\parallel$ and $\psi_\perp$, and then analogously to \eref{eq:ForceResponse} we would find that $\psi_\parallel = \nu_\parallel$ and $\psi_\perp = \nu_\perp$.
We have considered this previously \cite{Bradly2018}:
In the desorbed phase $T > \Ta$ the walks are free in the $d$-dimensional half-space and so clearly $\nu_\perp = \nu_\parallel = \nu_d$.
In the adsorbed phase $T < \Ta$ the fraction of the walk in the surface approaches unity and so $\nu_\perp = 0$ and $\nu_\parallel = \nu_{d-1}$.
At the adsorption point $T = \Ta$ we also find that $\nu_\perp = \nu_\parallel = \nu_d$ and in fact, this coincidence was a reliable method for locating the critical temperature.
The relevance to the current work is that the response to a small force near the adsorption point does not depend on the direction of that force, even when there is a surface.
Incidentally, $\mu$ is the same for the half-space and full lattices \cite{Hammersley1982}, whereas $\gamma$ does change if there is a surface \cite{Guttmann1984}.

\subsection{Phase re-entrance at low temperature}
\label{sec:Reentrance}

There is an argument, going back to Ref.~\cite{Orlandini2004} but also considered in \cite{Krawczyk2004, Osborn2010,Tabbara2012}, for understanding $T$--re-entrance by approximating the free energy of a polymer at low temperature using the zero-temperature contributions to energy and entropy $A \approx U(T = 0) - T S(T = 0)$.
At zero temperature the likely configuration in each phase depends on the angle $\theta$.
When $\theta = 90^\circ$, i.e. purely vertical pulling, the only configuration in the stretched phase is a taut vertical rod so there is no configurational entropy and the energy $U$ is just the work done to lift the endpoint to its maximum height, or $F$ per step.
In the adsorbed phase the endpoint is in the surface so there is no work done but every step feels the surface interaction and there is entropy due to configurations in the surface (if possible).
Thus for vertical pulling we have 
\begin{equation}
	A \approx n F - n - T n \log \mu_{d - 1}.
	\label{eq:FreeEnergyLowTVertical}
\end{equation}
where the second term is the interaction energy since $\epsilon = - 1$.

For $\theta < 90^\circ$ there is now a horizontal component to the force and so the zero temperature configuration in the adsorbed phase is a taut rod lying entirely in the surface.
There is no longer any configurational entropy for the part that is in the surface so the free energy in the adsorbed phase consists only of the horizontal work $F \cos \theta$ per step.
In the stretched phase, for very large $F$, the general configuration is a staircase walk with $m$ horizontal steps such that $\tan \theta = (n - m) / m$.
At high temperature all $\binom{n - m}{m}$ such walks may be equally likely, but at low temperature and as the force reduces to the critical force, the surface interaction becomes significant and so configurations with more horizontal steps in the surface are more likely, since $\epsilon < 0$.
Thus the only relevant configuration is an L-shape with the first $m$ steps forming a rigid rod in the surface and the remaining $n-m$ steps are vertical.
The free energy of this configuration is the interaction energy $-m$ for the part in the surface and the work done on each part in the horizontal and vertical directions, $m F \cos \theta$ and $(n - m) F \sin \theta$, respectively.
In fact the adsorbed phase is just the L-shaped configuration where $m = n$.
The total free energy at low temperature near the critical force for $\theta < 90^\circ$ may be written \cite{Tabbara2012}
\begin{equation}
	A \approx - m + m F \cos \theta + (n - m) F \sin \theta.
	\label{eq:FreeEnergyLowTGeneral}
\end{equation}
This can be minimised with respect to $m$ to obtain an expression for the critical force at low temperature.
Combining with the case $\theta = 90^\circ$ \eref{eq:FreeEnergyLowTVertical} for vertical pulling we have
\begin{equation}
	F_c = \begin{cases}
		T \log \mu_{d - 1} + 1 & \theta = 90^\circ, \\[1ex]
		\displaystyle \frac{1}{\sin \theta - \cos \theta} & 45^\circ < \theta < 90^\circ, \\[2ex]
		\text{N/A} & \theta \leq 45^\circ.
	\end{cases}
	\label{eq:CriticalForce}
\end{equation}
where the third case follows from the second, i.e.~there is no force-induced desorption when $\theta \leq 45^\circ$.

To explain the $T$--re-entrance of the desorbed phase, we see from the first case of \eref{eq:CriticalForce} that for vertical pulling the line of critical force has non-zero dependence on $T$ as $T \to 0$.
This only occurs for $d = 3$ where the entropy of the adsorbed phase is non-zero since $\mu_2 > 1$.
For $d = 2$ there is no $T$--re-entrance since $\mu_1 = 1$.
As the angle decreases from $90^\circ$, we see from the second case of \eref{eq:CriticalForce} that $F_c$ is independent of $T$ as $T \to 0$ but also that the value of $F_c$ at $T = 0$ increases. 
So the $T$--re-entrance observed in the phase diagrams \fref{fig:OrderParameters3D} `softens' as the angle is decreased, and must eventually disappear by $\theta = 45^\circ$ where there is no force-induced desorption.
It is not known from this argument whether there is an intermediate angle where there is strictly no $T$--re-entrance or if the range of fixed $F$ just becomes very small.

\begin{figure}
	\centering
	\includegraphics[width=0.5\textwidth]{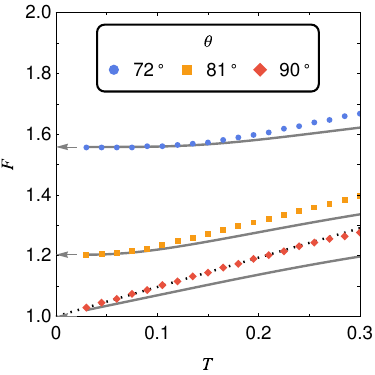}
	\caption{The phase boundaries for $d = 3$ walks at low temperature and near-vertical pulling.
	The points are local peaks of the Hessian eigenvalue $\log\chi$ from simulations of SAWs.
	The dotted black line has slope $\log\mu_2$ for comparison to the SAW boundary at $\theta = 90^\circ$.
	The gray lines are the boundaries of the PDW model and the gray arrows are the zero temperature values of the critical force from \eref{eq:CriticalForce}.}
	\label{fig:PhaseBoundariesLowT}
\end{figure}

While the above arguments are applicable for general lattice polymer models the low-temperature details have not been explored for SAWs specifically.
With additional simulations, we show in \fref{fig:PhaseBoundariesLowT} the phase boundaries for near-vertical pulling angles at low temperature for $d = 3$ SAWs.
The points indicate where there is a peak in the covariance measured by the Hessian eigenvalue $\log \chi$. 
The boundaries in \fref{fig:PhaseBoundariesLowT} clearly show the behaviour indicated by \eref{eq:CriticalForce} as $T \to 0$, while maintaining $T$--re-entrance as $T$ increases.
In particular, for $\theta = 90^\circ$ the phase boundary is linear in $T$ as $T \to 0$ with a slope given by the entropy of the adsorbed phase, namely $\log \mu_2$ which is the slope of the dotted black line.
Then, as $\theta$ decreases from vertical, there is a small region where the critical force is independent of temperature and this region expands as $\theta$ decreases.
The arrows indicate the $T = 0$ values of the critical force from \eref{eq:CriticalForce} and the boundaries are approaching these values.

In order to access temperatures as low as $T \geq 0.03$ it was necessary to restrict to length $n \leq 128$ for this range of $F$ and $\theta$.
It is reasonable to worry about finite-size effects in this regime, but in our experience this would be below the resolution of the parameter space \cite{Krawczyk2004}.
Lastly, the gray lines are the boundaries for the partially directed walk model at the same angles.
The comparison will be discussed more in the next section.

\subsection{Comparison to PDWs}
\label{sec:PDWComparison}

The phase diagram of SAWs is very similar to the partially directed walk models as illustrated by the schematic phase diagram in \fref{fig:Schematic}, with a few minor differences.
We first recapitulate some details of the partially directed walk (PDW) model for polymers pulled at an angle, following \cite{Osborn2010,Orlandini2010}.
For $d = 2$ the valid partially directed walks are a subset of self-avoiding walks with the restriction that steps in the negative $x$-direction are forbidden so that the directedness of the walk aligns with the horizontal component of the applied force.
PDW models can often be solved exactly whereby instead of a partition function the description begins with a generating function
\begin{equation}
	P(z, \kappa, \lambda, \tau) = \sum_{n,m,a,h} p_{nmah} z^n \kappa^m \lambda^a \tau^h,
	\label{eq:PDWGeneratingFunction}
\end{equation}
where $p_{nmah}$ is the number of PDWs of length $n$ with $m$ surface contacts and endpoint position $(a,h)$.
The weights $\kappa$, $\lambda$ and $\tau$ are the same as before and $z$ is a fugacity associated with the length.
To obtain the phase diagram for the PDW model a closed form of the generating function $P$ is found exactly, for example using a factorisation scheme \cite{Orlandini2010} or kernel method \cite{Osborn2010,Prodinger2004}.
Once $P$ is known exactly its singularity structure can be examined, and the dominant singularity determines the phase as a function of $\kappa$, $\lambda$ and $\tau$.
Where two or more singularities coincide indicates a phase transition given by the condition \cite{Osborn2010}
\begin{equation}
	\lambda = \frac{\kappa \ \tau \left( \kappa - \kappa \ \tau^2 - 1 \right)}{(\kappa - 1) \left[(\kappa - 1)^2 - \kappa^2 \tau^2 \right]},
	\label{eq:PDWBoundary}
\end{equation}
which is valid for $d = 2$.
This condition can be converted to physical parameters $T$, $F$ and $\theta$ using \eref{eq:BoltzmannToPhysical}.

For $d = 3$ there is flexibility in how the partially directed walk model is defined.
The simplest definition would be similar to the $d = 2$ definition and only prevent steps in the negative $x$-direction.
However, this model is not solvable, as it is equivalent to a sequence of two-dimensional self-avoiding walks in the $y$-$z$ plane.
Instead, by allowing positive and negative steps in the $z$ direction (perpendicular to the interacting surface) but only positive steps in the $x$ and $y$ directions, the model becomes solvable.
These $d = 3$ walks remain partially directed when projected down to the $x$-$y$ adsorbing surface. 
For this model the condition for the critical force changes to $\lambda + 1$ on the left-hand side of \eref{eq:PDWBoundary}, with the same right-hand side.
We use this model for comparison to SAWs.
Ref.~\cite{Osborn2010} also considered a second solvable model where $d = 3$ walks are generated from $\mathbb{Z}$-coloured $d = 2$ walks, but while this model has some quantitative differences, there is no new physics, so we do not consider it here.

In the limit of zero force, (equivalent to $\lambda \to 1$ and $\tau \to 1$) \eref{eq:PDWBoundary} indicates a thermal transition at the adsorption point $\Ta = 1 / \log(1 + 1/\sqrt{2}) = 1.87\ldots$ for $d = 2$ and $\Ta = 1 / \log(7/8 + \sqrt{17}/8) = 3.03\ldots$  for $d = 3$, independent of angle $\theta$.
Thus the adsorption point for PDWs is slightly higher for $d = 2$ and lower for $d = 3$.
However, we also note that the PDW models count edge-surface interactions where as our SAW model counts vertex-surface interactions.
Also in the $F \to 0$ limit it can be shown that there is an angle $\theta_* = \tan^{-1}1/2 \approx 27^\circ$ where the phase boundary is vertical.
A corollary of this feature is that for $\theta_* < \theta < 45^\circ$ there is a narrow temperature range whereby there is $F$--re-entrance.
That is, for such angles, and some $T$ just below $\Ta$ the model is in the adsorbed phase at low $F$ and as $F$ increases the model desorbs and then re-adsorbs at higher $F$.
However, even for the PDW model this feature is hard to distinguish and our numerical data is not good enough to resolve this feature in the SAW model.
However, the fact that the boundary varies smoothly with $\theta < 45^\circ$, indicates that some $\theta_*$ exists for the SAW model.
Based on the similarity between the two models, and comparison of the SAW boundaries for angles below $45^\circ$ suggest that it is possible for $\theta_*$ to be the same for both models.

The other region of interest is the limit of zero temperature, which is equivalent to the limit $\kappa \to \infty$ for \eref{eq:PDWBoundary}, whereby $\lambda \sim \tau/\kappa$.
After converting to physical variables there is a force-induced desorption transition at the critical value $F_c = 1/(\sin\theta \,-\, \cos\theta)$, precisely equivalent to the free energy argument of \eref{eq:CriticalForce} but with a different value for the adsorbed phase entropy.
Due to this similarity we conclude that low-temperature force-induced desorption is only possible in the SAW model for $\theta > 45^\circ$.
In other words, with respect to \fref{fig:PhaseBoundaries}, we expect that the phase boundary for $\theta = 45^\circ$ is asymptotically vertical for large $F$, and the phase boundary for $\theta = 45^\circ$ will intersect the $F$ axis, even though we cannot simulate at such large values of $F$.

As temperature increases from near zero, the $d = 3$ PDW model displays $T$--re-entrance and the $d = 2$ model does not, the same as the SAW model.
In \fref{fig:PhaseBoundariesLowT} we have also directly compared the solution for the PDW model to the SAW model in the low temperature regime via solid gray lines indicating solutions to \eref{eq:PDWBoundary}.
For vertical pulling $\theta = 90^\circ$ the difference in models is due to the difference in entropy of the adsorbed phases.
The slope of the PDW boundary in this case is $\log\left(1 + \sqrt{2}\right)$ which is less than the slope of the SAW boundary $\log\mu_2$, indicated by the dotted black line.
For angles $\theta < 90^\circ$, the difference in adsorbed phase entropy between the models also affects the $T$--re-entrant region.
Interestingly however, the region where the critical force is independent of temperature for $\theta < 90^\circ$ seems to be quite close for both the SAW and PDW models: it would be intriguing if the directed and non-directed models shared this value.

Ref.~\cite{Osborn2010} also considered that for $d = 3$ there is further flexibility in that the force can be applied such that the `horizontal' component is not parallel to the $x$-axis, but could be e.g.~along the $x = y$ line.
In this case there is still configurational entropy in the adsorbed part of the walk that ends at some point $(x,x,0)$.
This leads to a slight change in the low temperature part of the phase diagram due to the additional angle of the applied force relative to the lattice directions, namely the slope of the phase boundary remains non-zero as $T \to 0$.
However, because this is a lattice effect, and because our SAW model is already constrained by how we can simulate the SAWs, we have only considered the force to be in the $x$-$z$ plane for the $d = 3$ models.

\section{Conclusion}
\label{sec:Conclusion}

We have considered a self-avoiding walk model of lattice polymers pulled by a force that is applied at varying angles relative to an interacting surface.
Monte Carlo simulations were used to calculate order parameters and sketch the phase diagram.
We find that there are two phases, adsorbed and desorbed, characterised by the average fraction of the walk that is in contact with the surface.
Both phases exist for all angles but the location of the boundary between the phases is strongly dependent on the angle of the applied force.
For vertical and near-vertical pulling there is a critical force above which the walk is desorbed from the surface, and this only occurs for low temperatures; at high temperatures the walk is always desorbed.
Simple free energy arguments indicate that this is the case for any $\theta > 45^\circ$.
For horizontal and near-horizontal pulling the converse holds: at high temperatures there is a critical force below which the walk is desorbed but at low temperatures the walk is always adsorbed.
At intermediate angles there is a small range of temperature where increasing force can induce both desorption and adsorption.

We can compare our results to previous studies of partially directed walk models finding good qualitative agreement. 
Minor quantitative agreements are due to differences in the location of the adsorption transition and configurational entropy which affects the location of the critical force especially in the low temperature regime.

A remaining question concerns the difference in the phase transitions between the two models when the pulling is horizontal ($\theta=0^\circ$).
For PDWs horizontal pulling always produces a second-order transition since the horizontal span of PDWs in the adsorbed phase is always linear in length even at $F=0$.
Similarly, for SAWs in the adsorbed phase with non-zero horizontal force the scaling of the horizontal extension is always $\langle a \rangle \sim n$.
As $F \to 0^+$ we expect $\sqrt{\langle a ^2 \rangle} \sim n$ for $d=2$ SAWs since there are only two adsorbed configurations (and $\langle a \rangle$ is not meaningful in this case).
However, for $d = 3$ the adsorbed phase will have $\langle a \rangle \sim n^{\nu}$, where $\nu = 3/4$ is the metric exponent for two-dimensional SAWs.
This suggests a possible further transition as $F\to0^+$ but as mentioned around \fref{fig:PhaseBoundaries} our data contains spurious features due to the definition of the model.
The better analysis would consider the average of the squared end-to-end distance $R^2$, and focus on the transition, which could be the subject of future study with new simulations focusing on purely horizontal pulling.

Regarding experiments, the observed dependence of the critical force \cite{Grebikova2018} was performed for temperature just below $\Ta$. The measured desorption force agrees with the PDW model for near-vertical pulling, but increases more slowly than that model as the angle decreases. 
The desorption force in fact diverges in the PDW model as $\theta \to \theta_* \approx 27^\circ$ which explains why the gap with experimental data widens.
As for the SAW model, although we do not have excellent resolution in this region of the phase diagram, our data indicates that it is qualitatively the same as the PDW model, with the only difference being the value of $\Ta$.
Thus the difference to experimental data is best understood as a lattice effect.

Another experimental consideration is that the flat slope of the boundary for small but non-zero $T$ (and near-vertical pulling) was shown in the PDW model already \cite{Osborn2010,Orlandini2010}, but here we have demonstrated that it should be present in the full SAW model, which is canonically closer to physical polymers. 
This feature has not been observed in experiments although the range of small $T$ for which the critical force is independent of $T$ may be hard to access directly. 
However, this feature may affect the location of the critical force at higher $T$ where the desorbed phase is $T$--re-entrant.

\ack	
The authors acknowledge financial support from the Australian Research Council via its Discovery Projects scheme (DP230100674).
This research was supported by The University of Melbourne's Research Computing Services and the Petascale Campus Initiative.
Data generated for this study is available on request.

\section*{References}		
\bibliographystyle{plain}
\IfFileExists{../../../papers/bibtex/polymers_master.bib}
{\bibliography{../../../papers/bibtex/polymers_master.bib}}
{
	\bibliography{pulling-at-angle.bib}
}

\end{document}